\documentclass{aa}
\usepackage{graphicx}
\usepackage{natbib}
\usepackage{url}
\begin{document}
   \title{Experimental \ion{N}{v} and \ion{Ne}{viii} low-temperature dielectronic recombination rate coefficients}

   \author{S. B\"ohm\inst{1}\fnmsep\thanks{\email{Sebastian.G.Boehm@physik.uni-giessen.de}},
           A. M\"uller\inst{1}, S. Schippers\inst{1}, W. Shi\inst{1}, M. Fogle\inst{2}\fnmsep\thanks{Present Address: Physics Division, Oak Ridge
           National Laboratory, Oak Ridge, TN 37831-6372}, P. Glans\inst{2},
           R. Schuch\inst{2}, and
           H. Danared\inst{3}}

   \offprints{S. B\"ohm}

   \institute{Institut f\"ur Atom- und Molek\"ulphysik, Justus-Liebig-Universit\"at Giessen,
              Leihgesterner Weg 217, 35392 Giessen
         \and
             Atomic physics, Fysikum, Stockholm University, AlbaNova S-10691 Stockholm, Sweden
         \and
             Manne Siegbahn Laboratory, 10405 Stockholm, Sweden
             }

   \date{}

   \abstract{
   The dielectronic recombination rate coefficients of \ion{N}{v} and \ion{Ne}{viii} ions have been
   measured at a heavy-ion storage ring. The investigated
   energy ranges covered all dielectronic recombination resonances attached to $2s \rightarrow 2p$
   ($\Delta n=0$) core excitations. The rate coefficients in a
   plasma are derived and parameterized by using a convenient fit formula. The experimentally derived
   rate coefficients are compared with theoretical data by Colgan et al.\
   (2004, A\&A, 417, 1183) and Nahar~\&~Pradhan (1997, ApJ, 111, 339) as well as with the recommended rate coefficients
   by Mazzotta~et~al. (1998, A\&A, 133, 403). The data of
   Colgan~et~al. and Nahar \& Pradhan reproduce the experiment
   very well over the temperature ranges where \ion{N}{v} and
   \ion{Ne}{viii} are expected to exist in photoionized as well as
   in collisionally ionized plasmas. In contrast, the recommendation of Mazzotta~et~al. agrees with the
   experimental rate coefficient only in the temperature
   range of collisional ionization. At lower temperatures it deviates from the measured rate
   coefficient by orders of magnitude.
   In addition the influence of external electric
   fields with field strengths up to 1300~V/cm on the dielectronic recombination rate coefficient has been investigated.

   }

   \authorrunning{S.~B\"ohm {\it et al.}}
   \titlerunning{\ion{N}{v} and \ion{Ne}{viii} dielectronic recombination rate coefficients}
   \maketitle
%

\section{Introduction}

One important process that governs the charge state balance in a
plasma is dielectronic recombination (DR). Accordingly, DR rate
coefficients form a basic ingredient in plasma modeling codes that
are employed for the analysis of spectra obtained from
astrophysical observations \citep{Ferland2003a}. In order to be
able to infer a reliable description of the plasma properties,
such as element abundances and temperatures, from such
calculations, accurate rate coefficients for the basic atomic
collision processes in a plasma are required. To date, most DR
rate coefficients used for plasma modelling stem from theoretical
calculations. A recent compilation of DR data for dynamic finite
density plasmas can be found in the Atomic Data and Analysis
Structure Database (ADAS) which can be accessed under
\texttt{http://www-cfadc.phys.ornl.gov/data\_and\_codes}
\citep{Badnell2003a}. The resolution and precision of this data is
tuned to spectral analysis and are sufficient for prediction of
the DR contributions to individual spectral line emissivities. The
calculation of DR rate coefficients is a challenging task since an
infinite number of states is involved in this process.
Approximations and computational simplifications are applied in
order to make DR calculations tractable. Hence, experimental
benchmarks are required for testing and improving the theoretical
methods. Here we present experimentally derived \ion{N}{v} and
\ion{Ne}{viii} DR rate coefficients.

In the storage ring measurements the ensembles of colliding
particles have a well defined average relative velocity. As a
consequence, high resolution is obtained in the whole range of
accessible energies and hence, cross sections can be measured in
great detail. In contrast to that, convolution of the measured
data with the plasma temperature leads to broad smooth dependences
of plasma rate coefficients $\alpha(T)$ where most details in the
cross sections are washed out. Nevertheless, the details of the
cross sections at low energies have a strong influence on the size
of the plasma rate coefficient at low temperatures. It is
difficult to predict such low lying DR resonances with sufficient
accuracy by theoretical calculations. An example has been given
recently by \citet{Schippers2004c} who studied DR of berylliumlike
\ion{Mg}{ix}. This ion exhibits strong DR resonances at
electron-ion collision energies below 100~meV. All theoretical
calculations show good agreement with experiment at high
temperatures, but show significant discrepancies at low
temperatures.

Low-energy DR of lithiumlike ions can be represented as
\begin{eqnarray}
  e^- &+& {\rm A}^{q+}(1s^22s) \to {\rm A}^{(q-1)+}(1s^2 2p\,nl)
 \nonumber\\
 &\to&\left\{
 \begin{array}{ll}
  {\rm A}^{(q-1)+}(1s^2 2s\,nl)+ h\nu & \mbox{\rm (type I)}\\
  {\rm A}^{(q-1)+}(1s^2 2p\,n'l') + h\nu' & \mbox{\rm (type II)}.
  \end{array}
  \right.
  \label{eq:DRLilike}
\end{eqnarray}
The $2s \rightarrow 2p$ ($\Delta n = 0$) excitation energy is
$\approx 10$~eV for N$^{4+}$ and $\approx 16$~eV for Ne$^{7+}$,
respectively. DR resonances that are associated with higher
excitations such as $2s \rightarrow 3l$ ($\Delta n=1$) that occur
at higher energies have not been measured.

The DR rate coefficient is sensitive to the presence of
electromagnetic fields. DR in the presence of external
electromagnetic fields (DRF) was studied first theoretically by
\citet{Burgess1969}, \citet{Jacobs1976} and \citet{Jacobs1979}. It
was recognized that an external electric field mixes states with
different orbital quantum numbers and thereby changes the
autoionization rates that are relevant for the DR process. The
resulting enhancement of the DR cross section has been verified
experimentally for the first time by
\citet{Mueller1986,Mueller1987}. DRF experiments providing
evidence for the influence of additional magnetic fields as
predicted by \citet{Robicheaux1997} were carried out by
\citet{Bartsch1999a,Bartsch2000} and \citet{Boehm2001b}.

The present paper is organized as follows. The experimental
procedure is outlined in Sec.~\ref{Sec:Exp}. In
Sec.~\ref{Sec:results} the experimental results are presented and
compared with theoretical results in Sec.~\ref{Sec:compare}. The
influence of electromagnetic fields on the DR rate coefficient is
discussed in Sec.~\ref{Sec:field}. A summary is provided in
Sec.~\ref{Sec:conclusion}


\section{Experiment}\label{Sec:Exp}

The \ion{N}{v} and \ion{Ne}{viii} DR experiments were carried out
at the heavy-ion storage-ring {\sc cryring} of the Manne Siegbahn
Laboratory in Stockholm. The \element[4+][14]{N} and
\element[7+][20]{Ne} ions were produced in a cryogenic electron
beam ion source ({\sc crysis}), preaccelerated by a radio
frequency quadrupole structure to about 300~keV/u, injected into
the ring and accelerated to their final energy of a few MeV/u. The
ion beam was cooled by interaction with a magnetically guided
colinear beam of cold electrons in an electron cooler
\citep{Danared2000}. During the recombination measurements the
electron cooler was used as an electron target. The electron
energy was varied by changing the voltage at the cooler cathode in
order to obtain a DR spectrum \citet{Zong1998}. The experimental
center of mass energy range of $0<\hat{E}<11$~eV for
\element[4+][14]{N} and $0<\hat{E}<20$~eV for
\element[7+][20]{Ne}, respectively, covered all DR resonances due
to $2s\rightarrow 2p$ ($\Delta n=0$) excitations. The recombined
ions were detected with $100\%$ efficiency behind the first dipole
magnet following the cooler.

\begin{figure}
\centering
\includegraphics[width=\columnwidth]{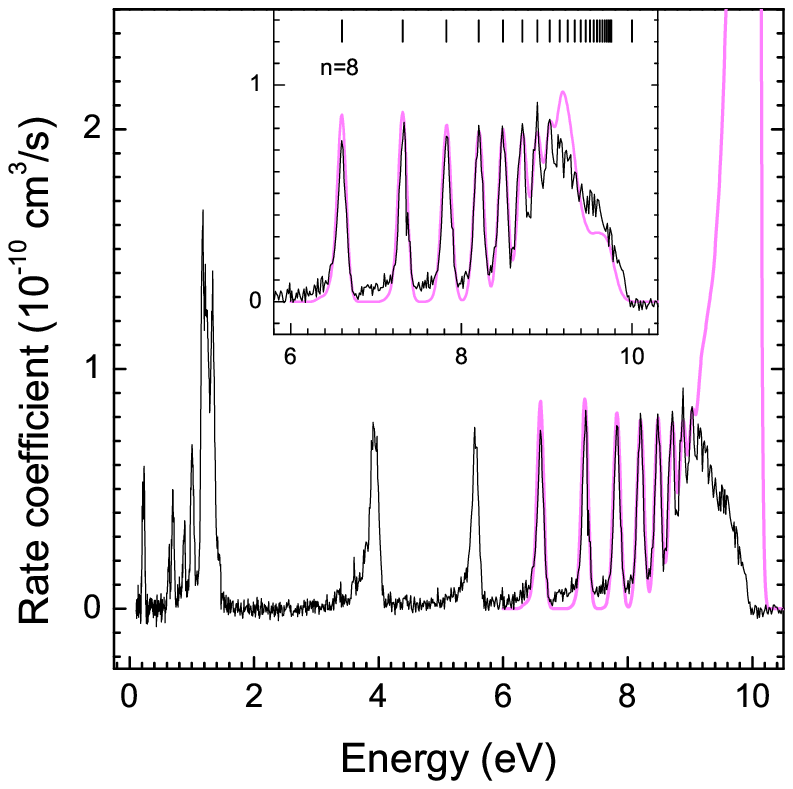}
\caption{Experimental zero-field N$^{4+}$ DR spectrum for an ion
energy of 7.5~MeV/u (thin solid line). The resonances are due to
the excitation of $1s^22p\,nl$ intermediate states with $n\geq 5$.
The thick line beginning at 6~eV ($n\geq 8$) shows the
extrapolation function obtained by adjusting an {\sc
autostructure} calculation to the experimental high-$n$ spectrum.
It includes Rydberg states up to $n=1000$. The inset shows the
high energy part of the experimental spectrum together with the
extrapolation function to which the field ionization model of
\citet{Schippers2001c} has been applied.} \label{Fig:N4_spe}
\end{figure}

On their way to the detector the recombined ions had to pass
strong magnetic fields. The largest field was that of the
charge-state analyzing bending dipole magnet. These magnetic
fields were perpendicular to the ions' flight direction and caused
motional electric fields (of the order of 50~MV/m in the ring
dipoles) orders of magnitude larger than any motional electric
fields in the interaction region. Since a sizeable fraction of
recombined ions was expected to be formed in highly excited and,
hence, very fragile states, the survival probability of these
states and the efficiency of their detection need particular
consideration. Recombined ions that reached a zone of large
motional electric fields while being in sufficiently highly
excited $1s^22p\,nl$ states were field ionized and therefore not
detected. This effect can be clearly seen in Fig.~\ref{Fig:N4_spe}
where the experimental \ion{N}{v} spectrum is compared with two
different calculations, one including basically all Rydberg states
that contribute to the DR cross section and one where the
field-ionization cut-off caused by the experimental conditions has
been modeled. The straight field ionization cut-off quantum-number
resulting from the simple over-the-barrier treatment is
$n_c\approx 17$. The detailed field ionization model
\citep{Mueller1987,Schippers2001c} accounts for the radiative
decay of electrons in high Rydberg states on their way from the
cooler to the different field ionization zones. The model uses
hydrogenic decay probabilities and field ionization rates in a
hydrogenic approach for all the individual Stark states developing
in the external fields. Compared with a simple step function at
$n_c = 17$ the model provides a realistic survival pattern of DR
contributions from a band of Rydberg states around $n_c$. From the
good agreement of the model calculations with the experiment
(insets of Fig.~\ref{Fig:N4_spe} and ~\ref{Fig:Ne7_spe}) we
conclude, that field ionization is sufficiently well understood in
our experimental setup.

For the DRF measurements electric fields were introduced in the
interaction region with the aid of magnet coils mounted inside the
main solenoid of the cooler for field corrections. The solenoid
produced the longitudinal field $B_{\|}$ needed to guide the
electrons through the cooler. The correction coils usually serve
for optimizing the alignment of the electron beam with respect to
the ion beam. In the present experiments they were used to
introduce well defined transverse magnetic field components
$B_{\bot}$ which transformed to electric fields
$E_{\bot}=v_iB_{\bot}$ in the rest frame of an ion moving with
velocity $v_i$. In order to avoid confusion it should be noted
that the motional electric field $E_{\bot}$ in the interaction
region, which caused the enhancement of the DR cross section, was
different and spatially well separated from the motional electric
fields $F$ which caused field ionization of recombined ions in
high Rydberg states. More comprehensive descriptions of the DRF
experiments at {\sc cryring} were given by
\citet{Boehm2001b,Boehm2002a}.

The main uncertainties of the measured rate coefficient arise from
the measurement of the ion current with a current transformer
($\approx 10{\%}$) and the uncertainty of the interaction length
($\approx 5{\%}$). All uncertainties add up to $\pm 15\%$
uncertainty of the absolute rate coefficients obtained with the
narrow electron energy distribution of the storage-ring
experiment.


\section{Results}\label{Sec:results}

\begin{figure}
\centering
\includegraphics[width=\columnwidth]{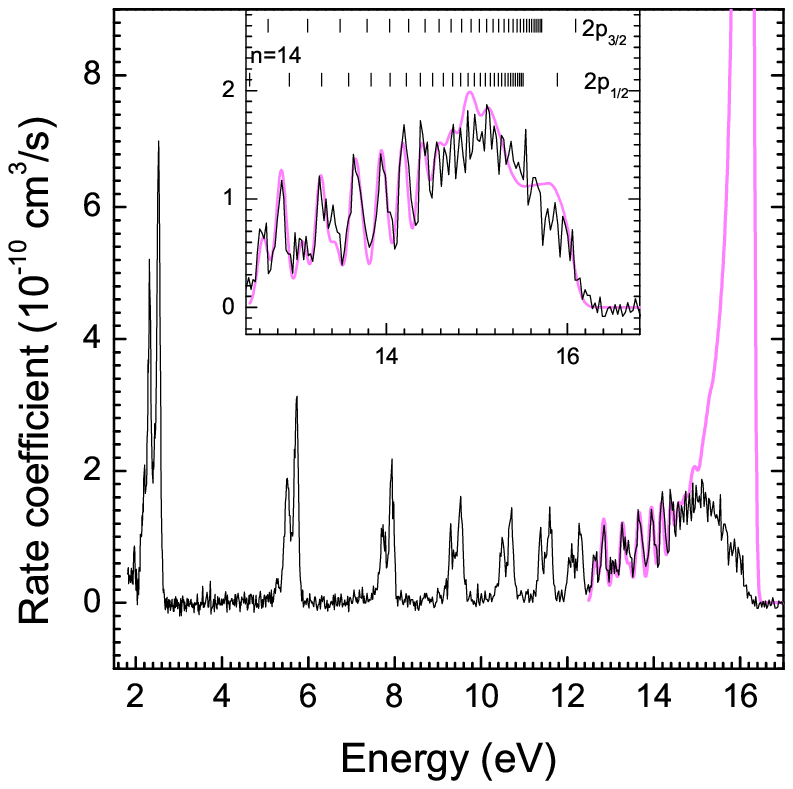}
\caption{Experimental zero-field Ne$^{7+}$ DR spectrum for an ion
energy of 11.4~MeV/u (thin solid line). The resonances are due to
the excitation of $1s^22p\,nl$ intermediate states with $n\geq 7$.
The thick line beginning at 12.45~eV ($n\geq 14$)shows the
extrapolation function obtained by adjusting an {\sc
autostructure} calculation to the experimental high-$n$ spectrum.
It includes Rydberg states up to $n=1000$. The inset shows the
high energy part of the experimental spectrum together with the
extrapolation to which the field ionization model of
\citet{Schippers2001c} has been applied.} \label{Fig:Ne7_spe}
\end{figure}

Our experimental \ion{N}{v} and \ion{Ne}{viii} DR rate
coefficients $\alpha(\hat{E})$ are displayed in
Fig.~\ref{Fig:N4_spe} and Fig.~\ref{Fig:Ne7_spe}, respectively.
They contain all $1s^22p\,nl$ resonances associated with $2s
\rightarrow 2p$ transitions up to $n \approx n_c$ ($n_c\approx 17$
for \ion{N}{v} and $n_c\approx 28$ for \ion{Ne}{viii}). For the
lowest $n$, even the $l$ substates are partially resolved.
Background subtraction (here radiative recombination is regarded
as background) was achieved by fitting an empirical formula to
those parts of the measured spectra where no DR resonances
occurred (Eq.~2 in \citet{Boehm2003a}).

For the derivation of a meaningful plasma rate coefficient from
our experimental data we have to estimate how much DR strength was
not observable in the experiment due to the field ionization
cut-off discussed above. Following the procedure described by
\citet{Schippers2001c} we have extrapolated the measured DR
spectrum to high-$n$ Rydberg states. We take advantage of the fact
that for these high Rydberg states, which are to be restored by
the extrapolation, simple scaling laws can be used to calculate
the DR cross section. The {\sc autostructure}  code
\citep{Badnell1986} is a convenient tool to make such an
extrapolation. The {\sc autostructure} extrapolation function
including Rydberg states up to $n=1000$ (states with $n>1000$ do
not contribute significantly anymore) is matched to the high-n
Rydberg region of the experimental spectrum  by applying a
constant energy shift of 0.05~eV towards higher energies and a
scaling factor of 1.15 to the calculated DR cross section for
\ion{N}{v}. The resulting function which is based on the {\sc
autostructure} calculation and which is used for extrapolating the
experimental rate coefficients is shown in Fig.~\ref{Fig:N4_spe}
from the region just above $n = 8$ to the high-n Rydberg
contributions together with the experimental data. (It should be
noticed that we do not intend to compare theory to experiment in
this context. This will be done later in this paper in connection
with the discussion of plasma rate coefficients.) For
\ion{Ne}{viii} the matching between experiment and extrapolation
function was achieved for $n\geq 14$ by applying a correction
factor of 1.63 to the radiative rate of the inner shell
transitions $2p_j \rightarrow 2s_{1/2}$ and an energy shift of
0.04~eV towards higher energies. Again, this manipulation does not
imply an attempt to correct the {\sc autostructure} calculation,
it is just meant as a way to adjust a meaningful function to the
experimental data. The function is then used for the required
extrapolation. The result obtained for the extrapolation function
is shown in Fig.~\ref{Fig:Ne7_spe}. While the measured rate
coefficients have an uncertainty of $\pm 15\%$ the possible error
of the extrapolated cross sections is difficult to quantify. We
estimate an uncertainty of $25 \%$ for the combined error of the
resulting plasma rate coefficients.

Plasma rate coefficients have been derived by convoluting the
experimental DR cross section $\sigma(\hat{E})=\alpha(\hat{E})
\times \sqrt{m_e/2\hat{E}}$ including the high-$n$ extrapolation
with a Maxwellian electron energy distribution yielding
\begin{eqnarray}
\alpha(T_{\rm e}) &=& (k_{\rm B}T_{\rm e})^{-3/2}\frac{4}{\sqrt{2
m_{\rm e}\pi}}\nonumber\\ &&\times\int_0^\infty d\hat{E} \,
\sigma(\hat{E}) \hat{E} \exp{(-\hat{E}/k_{\rm B}T_{\rm e})}
\label{eq:alphaTsigma}
\end{eqnarray}
with the plasma electron temperature $T_{\rm e}$, the electron
rest mass $m_e$, and Boltzmann's constant $k_{\rm B}$. This
procedure is applicable as long as the relative energy $\hat{E}$
is larger than the experimental energy spread , i.e., for $T_{\rm
e} \gg 70$~K in the present case.

\begin{figure}
\centering
\includegraphics[width=\columnwidth]{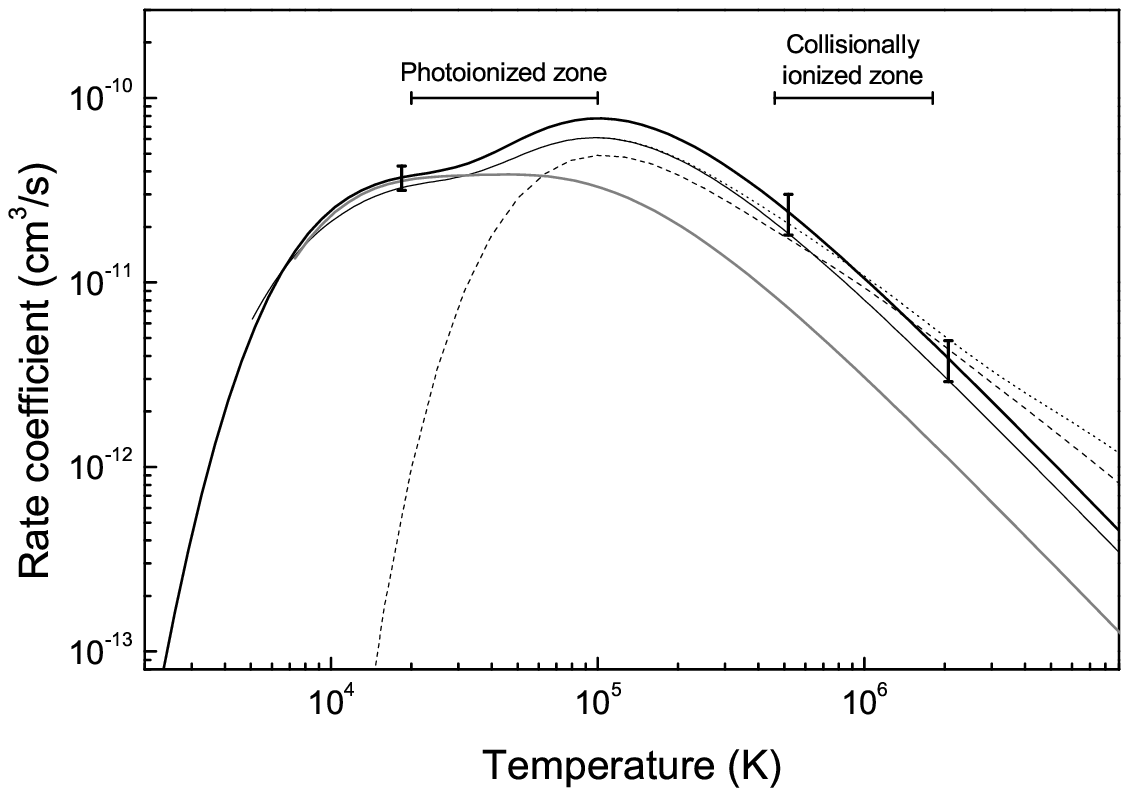}
\caption{The experimental \ion{Ne}{viii} $\Delta n=0$ DR rate
coefficient in a plasma is shown (thick solid line) with a
systematic uncertainty of $\pm 15\%$ below 40\,000~K and $\pm
25\%$ above. The thick grey line represents our measured rate
coefficient without the {\sc autostructure} extrapolation (see
text). The theoretical rate coefficient as calculated by
\cite{Colgan2004a} is shown for $\Delta n=0$~DR alone (thin solid
line) and with inclusion of $\Delta n=1$~DR (dotted line). Also
included is the theory-based rate coefficient inferred by
\citet[][ dashed line]{Mazzotta1998}.} \label{Fig:Ne7_ratecoeff}
\end{figure}

The experimentally derived DR rate coefficients with (thick black
line) and without (thick grey line) the extrapolation are shown in
Fig.~\ref{Fig:N4_ratecoeff} for \ion{N}{v} and in
Fig.~\ref{Fig:Ne7_ratecoeff} for \ion{Ne}{viii}. Above
$\approx20\,000$~K the rate coefficient for \ion{N}{v} and
\ion{Ne}{viii} is significantly influenced by the contributions of
high Rydberg states $n>17$ and $n>28$, respectively, restored by
the extrapolation.

A convenient representation of the plasma DR rate coefficient is
provided by the following fit formula
\begin{equation}
\alpha(T_{\rm e}) = T_{\rm e}^{-3/2} \sum_i c_i\exp{(-E_i/k_{\rm
B}T_{\rm e})}. \label{eq:alphafit}
\end{equation}
It has the same functional dependence on the plasma electron
temperature as the \citet{Burgess1965} formula, where the
coefficients $c_i$ and $E_i$ are related to oscillator strengths
and excitation energies, respectively. The results for the fit to
the experimental \ion{N}{v} $\Delta n=0$ DR rate coefficient in a
plasma are summarized in Table~\ref{tab:Nfit}. The fit deviates
from the thick full line in Fig.~\ref{Fig:N4_ratecoeff} by no more
than $0.2\%$ for $T_e\geq 5000$~K and by no more than $1\%$ for
2500~K~$\leq T_e <$~5000~K. Below 2500~K the DR rate coefficient
decreases rapidly and radiative recombination dominates the
recombination rate coefficient. The result for \ion{Ne}{viii} is
given in Table~\ref{tab:Nefit}. The fit deviates from the thick
full line in Fig.~\ref{Fig:Ne7_ratecoeff} by no more than $0.3\%$
for $T_e\geq 11\,000$~K and by no more than $2\%$ for 3000~K~$\leq
T_e <$11\,000~K.


\section{Comparison with theoretical results}\label{Sec:compare}

\subsection{\ion{N}{v}}

\begin{figure}
\centering
\includegraphics[width=\columnwidth]{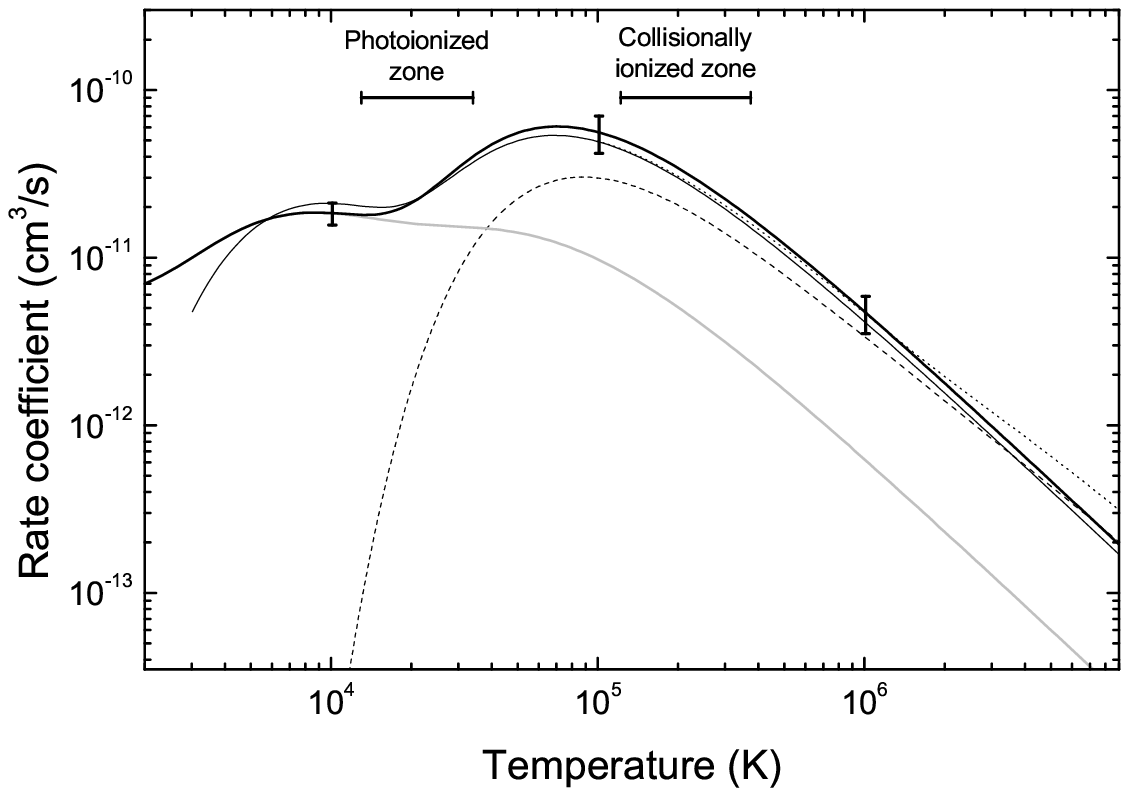}
\caption{The experimental \ion{N}{v} $\Delta n=0$ DR rate
coefficient in a plasma is shown (thick solid line) with an
systematic uncertainty of $\pm 15\%$ below 20\,000~K and $\pm
25\%$ above. The thick grey line represents our measured rate
coefficient without the {\sc autostructure} extrapolation (see
text). The theoretical rate coefficient as calculated by
\cite{Colgan2004a} is shown for $\Delta n=0$~DR alone (thin solid
line) and with inclusion of $\Delta n=1$~DR (dotted line). Also
included is the theoretical rate coefficient of \citet[][ dashed
line]{Mazzotta1998}.} \label{Fig:N4_ratecoeff}
\end{figure}

Our experimental \ion{N}{v} DR plasma rate coefficient is compared
with theoretical results in Fig.~\ref{Fig:N4_ratecoeff}. The
theoretical calculation of \cite{Colgan2004a} reproduces the
experimental result very well down to about 5000~K. This is well
below the temperature range where \ion{N}{v} is expected to form
in photoionized or in a collisionally ionized plasma. These
temperature ranges are also indicated in
Fig.~\ref{Fig:N4_ratecoeff}. They were estimated from the model
calculations of \citet{Kallman2001} as described by
\citet{Schippers2004c}. Also included is the result of
\cite{Colgan2004a} obtained with the inclusion of $\Delta n=1$~DR
which gives a significant contribution above 50\,000~K.

\begin{figure}
\centering
\includegraphics[width=\columnwidth]{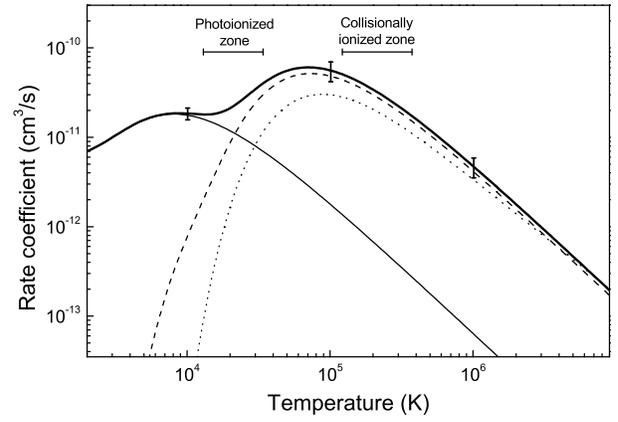}
\caption{The experimental \ion{N}{v} $\Delta n=0$ DR rate
coefficient in a plasma (thick solid line) is shown (including the
{\sc autostructure} extrapolation) with an systematic uncertainty
of $\pm 15\%$ below 20\,000~K and $\pm 25\%$ above. The thin solid
line shows our measured rate coefficient for the $n=5$ DR
resonance only. The dashed line shows our measured rate
coefficient excluding the $n=5$ DR resonance. Also included is the
theoretical rate coefficient of \citet[][ dotted
line]{Mazzotta1998}.} \label{Fig:N4_n5}
\end{figure}

\citet{Mazzotta1998} compiled DR rate coefficients and condensed
this compilation into a set of recommended DR rate coefficients.
They adopted the DR calculations of \citet{Chen1991} for the
lithiumlike ions who calculated total DR rate coefficients for 11
ions (Carbon, Oxygen, Neon, ..., Xenon). For the remaining ions,
such as \ion{N}{v}, \citet{Mazzotta1998} interpolated along the
iso-electronic sequence. This result is also included in
Fig.~\ref{Fig:N4_ratecoeff}. It is close to the experimental
result above $2 \times 10^5$~K. However, it has to be pointed out,
that \citet{Mazzotta1998} included $\Delta n \geq 1$~DR which has
not been measured in the present experiment. At lower temperatures
the plasma rate coefficient of \citet{Mazzotta1998} deviates
strongly from the present experimental result. The deviation can
be almost removed by excluding the $n=5$ DR resonance from the
experimental data as can be seen in Fig.~\ref{Fig:N4_n5}. This
leads us to the conclusion, that the result of
\citet{Mazzotta1998} misses to include the $n=5$ DR resonance
located between $\approx 0-1.5$~eV. The interpolation along
isoelectronic sequences seems to give sufficiently good results at
high temperatures but can deviate by orders of magnitude at low
temperatures due to the uncertainty in the inclusion of the
lowest-energy resonance contributions. The actual positions of
resonances at small energies strongly influence the plasma rate
coefficients at low temperatures.

\subsection{\ion{Ne}{viii}}

The experimental \ion{Ne}{viii} plasma rate coefficient is
compared with theoretical results in Fig.~\ref{Fig:Ne7_ratecoeff}.
The data of \cite{Colgan2004a} reproduces the experimental result
very well over the entire relevant temperature range. Also shown
in Fig.~\ref{Fig:Ne7_ratecoeff} is the result of
\cite{Colgan2004a} including $\Delta n=1$~DR which gives a
significant contribution above 300\,000~K. The recommended rate
coefficient of \citet{Mazzotta1998} reproduces the experimental
data well beyond $10^5$~K. Above $1.2 \times 10^6$~K the result of
\citet{Mazzotta1998} is above the experimental result. This is due
to the fact, that \citet{Mazzotta1998} included $\Delta n \geq
1$~DR which has not been measured in the present experiment. At
temperatures where \ion{Ne}{viii} is expected to form in
photoionized plasmas the recommendation of \citet{Mazzotta1998}
deviates from the experimental result by orders of magnitude. A
plausible reason for this is again in the sensitivity of
low-temperature plasma rate coefficients to the exact resonance
positions at low electron-ion energies.

\subsection{Total recombination rate coefficient}

\begin{figure}
\centering
\includegraphics[width=\columnwidth]{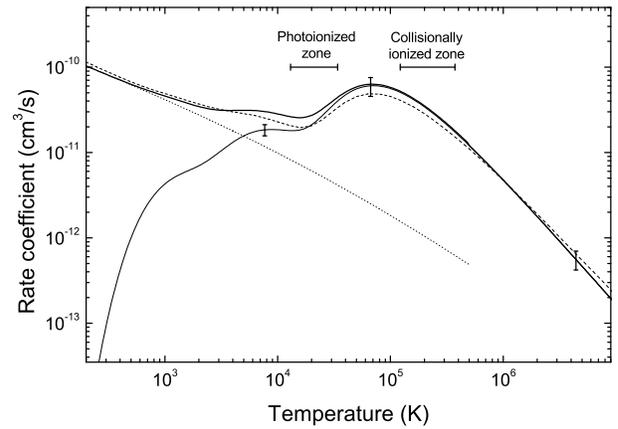}
\caption{Total \ion{N}{v} recombination rate coefficients in a
plasma: this work (thick full line, systematic error $\pm 15\%$
below 20\,000~K and $\pm 25\%$ above) and theoretical unified
calculation of \citet[][ dashed line]{Nahar1997}. Our total
recombination rate coefficient is obtained as the sum of the RR
rate coefficient of \citet[][ dotted line]{Pequignot1991} and our
DR rate coefficient (thin full line).} \label{Fig:N4_Nahar}
\end{figure}

A unified treatment of total recombination including DR and
radiative recombination (RR) has been performed by
\cite{Nahar1997}. Such a calculation in principle also accounts
for interference between RR and DR. In order to compare these data
with our experiment (see Fig.~\ref{Fig:N4_Nahar}) we have added
the theoretical RR rate coefficient by \cite{Pequignot1991} to our
experimental DR rate coefficient. The maximum deviation of theory
from experiment is less than $30\%$. It is found at temperatures
near $2\times 10^4$~K emphasizing again the crucial role of exact
resonance positions at low energies. The calculation of
\cite{Nahar1997} includes $\Delta n \geq 1$ transitions which is
the reason for the slight difference between theory and experiment
at high temperatures.

\begin{figure}
\centering
\includegraphics[width=\columnwidth]{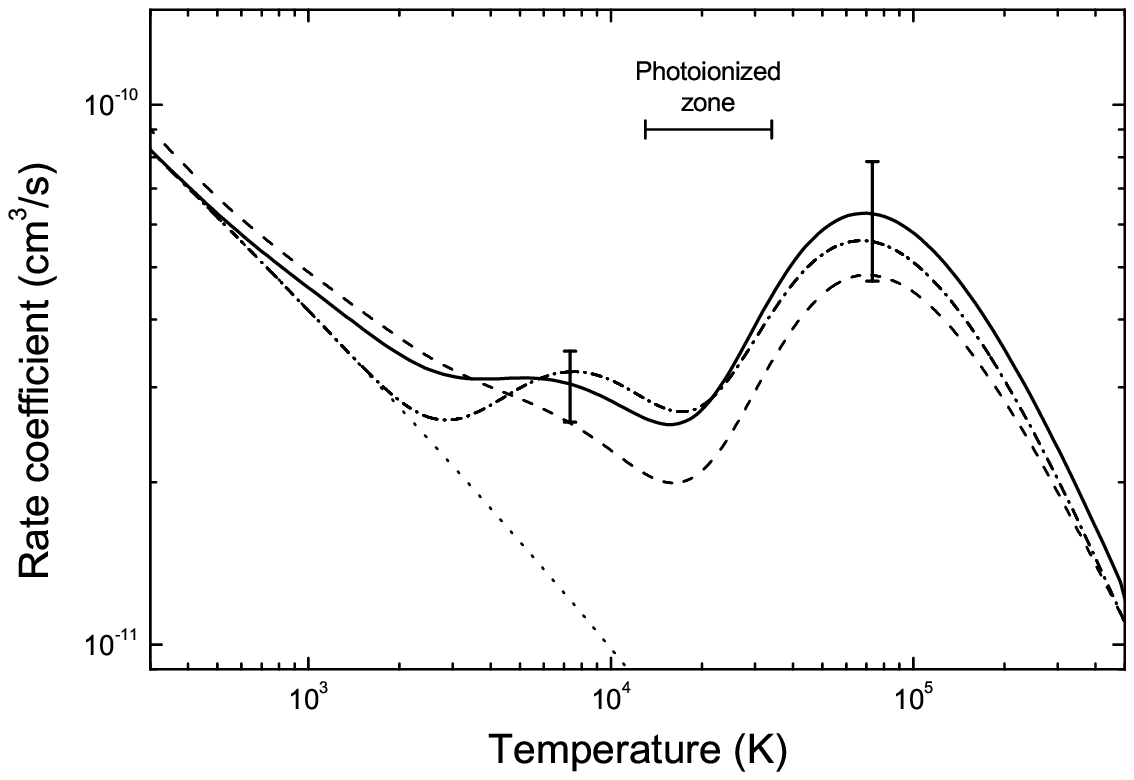}
\caption{Total \ion{N}{v} recombination rate coefficients in a
plasma: this work (thick full line, systematic error $\pm 15\%$
below 20\,000~K and $\pm 25\%$ above), the calculation by
\cite{Colgan2004a}  (dash-dotted line), and the unified
theoretical calculation of \citet[][ dashed line]{Nahar1997}. The
total recombination rate coefficients based on the experiment and
on the \cite{Colgan2004a} calculation are obtained by just adding
the RR rate coefficient as obtained by \citet[][ dotted
line]{Pequignot1991} to the DR data displayed in the previous
figures.} \label{Fig:N4_Nahar_enlarged}
\end{figure}

Fig.~\ref{Fig:N4_Nahar_enlarged} shows an enlarged part of the
plasma rate coefficient for \ion{N}{v}. Additional to the data
presented in Fig.~\ref{Fig:N4_Nahar} the calculation of
\cite{Colgan2004a} is now included to which the RR rate
coefficient by \cite{Pequignot1991} has been added. The agreement
with the experiment is well inside the experimental uncertainty
although this calculation does not include interference effects at
all. This finding together with the observed agreement between the
different theoretical approaches suggests that similar to DR of
\ion{C}{iv} ions \citep{Schippers2001c} and \ion{O}{vi} ions
\citep{Boehm2003a} interference effects do not play a significant
role in the recombination of \ion{N}{v} ions, either.


\section{Field effects}\label{Sec:field}

The DR rate coefficient can be strongly influenced by the presence
of electromagnetic fields. On the one hand electric fields ionize
electrons in high Rydberg states which reduces the recombination
rate coefficient. On the other hand electric fields mix
$l$-substates and thereby increase the recombination rate
coefficient \citep{Jacobs1976}.

In our experimental setup there is always an additional magnetic
field perpendicular to the electric field in the interaction
region which further influences the recombination rate coefficient
\citep{Robicheaux1997,Bartsch1999a} but has not been considered in
the theoretical investigations addressing the ions discussed in
this paper.

The present study of field effects on DR is relevant to low
density plasmas subject to large external fields e.\,g. solar
flares. The other major concern in connection with the role of
fields on DR in plasmas is that of the intrinsic plasma
microfield. Here the effect is density dependent; at high plasma
densities the rate enhancement due to field mixing is reduced by
continuum lowering, i.\,e., by collisions driving high-$n$ states
into local thermal equilibrium (LTE). Such conditions suppress the
high-$n$ states, that would otherwise contribute to enhanced DR
\citep[see][]{Badnell1993}. Generally as the uncertainties in the
basic atomic structure decrease the error in neglecting field
effects becomes more significant.

\begin{figure}
\centering
\includegraphics[width=\columnwidth]{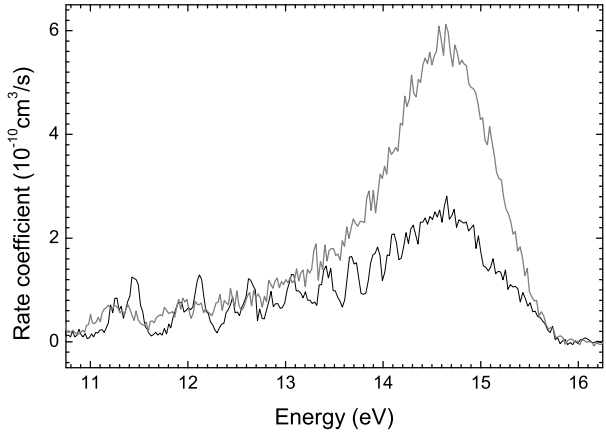}
\caption{\ion{Ne}{viii} DR spectra are shown for the field free
case (black solid line) and for the case of an applied electric
field of 1300~V/cm (thick grey line). The longitudinal guiding
field was 180~mT in both cases.} \label{Fig:Ne7_field_spe}
\end{figure}

\begin{figure}
\centering
\includegraphics[width=\columnwidth]{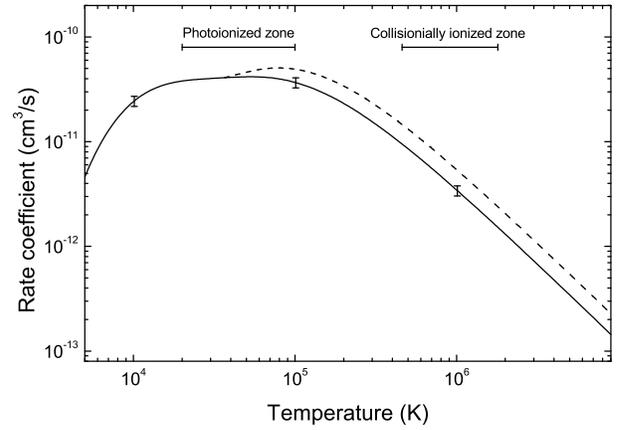}
\caption{Experimental \ion{Ne}{viii} DR plasma rate coefficient
(ion energy of
   11.4~MeV/u; including Rydberg states up to $n_c=28$) for the field free case
   (solid line) and in the presence of an
   electric field of 1300~V/cm (dashed line).} \label{Fig:Ne7_Efield}
\end{figure}

\ion{Ne}{viii} DR spectra are shown in
Fig.~\ref{Fig:Ne7_field_spe} for the field free case and for the
case of an externally applied electric field of 1300~V/cm. In both
cases the longitudinal magnetic guiding field was 180~mT. The
large increase of the recombination rate coefficient for high
Rydberg states is due to the electric field present in the
interaction region which is perpendicular to the longitudinal
magnetic guiding field and the ions' direction of flight. The
cut-off quantum number $n_c$ is the same for both spectra
($n_c=28$). The observed enhancement is a factor of 2.
Unfortunately the number of Rydberg states which could be measured
was limited by field ionization as described in
Sec.~\ref{Sec:Exp}. An extrapolation of the measured rate
coefficient in the presence of electromagnetic fields is not
possible at this time since the most advanced calculations
\citep{Griffin1998b,Griffin1998a} reproduce the experimental
result only qualitatively. For higher Rydberg states the
enhancement due to electromagnetic fields in the interaction
region grows as has been shown by $n$-differential measurements
\citep{Boehm2003b, Boehm2002a}. The enhancement of the DR rate
coefficient due to pure electric fields of \ion{Ne}{viii} has been
calculated for individual Rydberg states from $n=10-40$
\citep{Griffin1998b}. For an electric field of 100~V/cm the
enhancement factor increases from $\approx 1$ at $n=10$ to
$\approx 5$ at $n=40$. This shows that under certain circumstances
it is essential to account for the effect of electromagnetic
fields on the DR rate coefficient. The experimental plasma rate
coefficient with and without fields for \ion{Ne}{viii} is shown in
Fig.~\ref{Fig:Ne7_Efield}. Here the maximum enhancement observed
is $60\%$ at an electric field of $E=1300$~V/cm and a magnetic
field of $B=180$~mT. The cut-off quantum number was $n_c=28$. The
measured enhancement of $60 \%$ is already well beyond the
deviation of experiment and theoretical calculations as described
above.


\section{Conclusions}\label{Sec:conclusion}

The $\Delta n=0$~DR rate coefficients of \ion{N}{v} and
\ion{Ne}{viii} have been measured for $n<17$ and $n<28$,
respectively. After extrapolation to $n=1000$, $\Delta n=0$ DR
plasma rate coefficients were obtained from the data. The results
were compared with theoretical plasma rate coefficients of
\cite{Colgan2004a} and \citet{Nahar1997} as well as with the
recommended rate coefficient by \cite{Mazzotta1998}. The results
of Colgan~et~al. reproduce the experimental results very well for
both ions even at low temperatures. The data recommended by
Mazzotta~et~al. do not reproduce the experimental results for
temperatures below $10^6$~K. This is a consequence of the
sensitivity of low-temperature plasma rate coefficients to the
exact position of low energy resonances, which were not accessible
to \cite{Mazzotta1998}.

The deviation can be almost removed by excluding the $n=5$ DR
resonance from the experimental data. This leads us to the
conclusion, that the result of \citet{Mazzotta1998} misses to
include the $n=5$ DR resonance located between $\approx 0-1.5$~eV.
The interpolation along isoelectronic sequences seems to give
sufficiently good results at high temperatures but can deviate by
orders of magnitude at low temperatures due to the uncertainty in
the inclusion of the lowest-energy resonance contributions.

The influence of electric and magnetic fields on DR has been
investigated experimentally. A significant enhancement of the DR
rate coefficient was observed. A direct comparison with existing
theoretical data is not possible because present-day theory does
not include the magnetic field which was present in the
experiments besides perpendicular electric field components.

Our results bear important implications for the modelling of
cosmic plasmas where external fields are ubiquitous
\citep{Widrow2002}. In solar flares, e.~g., electric fields up to
1.3~kV/cm have been observed \citep{Zhang1986}.

\begin{acknowledgements}

We thank N.~R.~Badnell for helpful discussions and N.~Ekl\"ow,
G.~H~Dunn, N.~Djuri\'c, B.~Jelenkovi\'c, K.~MacAdam, S.~Madzunkov,
and W.~Zong for support during the experiment. This work was
funded by Deutsche Forschungsgemeinschaft under contract Schi
378/5-1 and M\"u 1068/8-1/2 and by the German Federal Ministry of
Education and Research under contract 06GI 148.

\end{acknowledgements}

\bibliographystyle{aa}
\bibliography{e:/tex_zeug/k3}
\newpage

\begin{table}
\centering
 \caption{ Fit parameters of the experimentally inferred \ion{N}{v} $\Delta n=0$ DR rate coefficient using
 Eq.~(\protect\ref{eq:alphafit}). Units are cm$^3$s$^{-1}$K$^{1.5}$ for $c_i$, and eV for $E_i$. The systematic
error of the rate coefficient $\alpha(T_e)$ from
Eq.~(\protect\ref{eq:alphafit}) is $\pm 15\%$.\label{tab:Nfit}}
 \begin{tabular}{ccc}
   $i$ & $c_i$ & $E_i$ \\
  \hline
  1....  & 2.410E-3 & 9.907 \\
  2....  & 1.710E-3 & 9.734 \\
  3....  & 2.835E-4 & 11.11 \\
  4....  & 7.197E-4 & 9.287 \\
  5....  & 1.356E-4 & 4.712 \\
  6....  & 6.171E-5 & 1.180 \\
  7....  & 2.945E-6 & 0.2815 \\
\hline
 \end{tabular}
\end{table}

\begin{table}
\centering
 \caption{ Fit parameters of the experimentally inferred \ion{Ne}{viii} $\Delta n=0$ DR rate coefficient using
 Eq.~(\protect\ref{eq:alphafit}).
 Units are cm$^3$s$^{-1}$K$^{1.5}$ for $c_i$, and eV for $E_i$. The systematic error of the rate coefficient
$\alpha(T_e)$ from Eq.~(\protect\ref{eq:alphafit}) is $\pm
15\%$.\label{tab:Nefit}}
 \begin{tabular}{ccc}
   $i$ & $c_i$ & $E_i$ \\
  \hline
  1....  & 5.992E-3 & 16.41 \\
  2....  & 5.492E-3 & 14.68 \\
  3....  & 6.338E-4 & 6.714 \\
  4....  & 3.394E-4 & 2.354 \\
  5....  & 8.214E-6 & 1.601 \\
  6....  & 7.205E-6 & 1.584 \\
  7....  & 2.122E-6 & 82.86 \\
\hline
 \end{tabular}
\end{table}

\clearpage

\end{document}